\documentclass[reprint,twocolumn,superscriptaddress,amsmath,amssymb,nofootinbib,preprintnumbers]{revtex4-2} 

\usepackage{mathptmx}
\usepackage{graphicx}  
\usepackage[mathlines]{lineno}
\usepackage[english]{babel} 
\usepackage{xcolor}
\usepackage{ulem}
\usepackage{dcolumn}   
\usepackage{bm}        
\usepackage{amssymb}   
\usepackage{amsbsy}   
\usepackage{comment}
\usepackage{multirow}
\usepackage{diagbox}
\usepackage[draft]{todonotes}
\usepackage{mathtools}

\usepackage{hyperref}
\hypersetup{
    colorlinks=true,
    linkcolor=cyan,
    filecolor=magenta,      
    urlcolor=cyan,
    citecolor=violet,
}

\makeatletter
\g@addto@macro\bfseries{\boldmath}
\makeatother

\def \LambdaI {\Lambda_\text{I}}
\def \Trh {T_\text{rh}}
\def \Hinf {H_\text{inf}}
\def \MPl {M_\text{Pl}}
\def \nn {\mathbf{n}}
\def \dif {\mathrm{d}}

\begin{document}

\title{Cosmological implications of photon-flux upper limits at ultra-high energies in scenarios of Planckian-interacting massive particles for dark matter}

\author{P.~Abreu}
\affiliation{Laborat\'orio de Instrumenta\c{c}\~ao e F\'\i{}sica Experimental de Part\'\i{}culas -- LIP and Instituto Superior T\'ecnico -- IST, Universidade de Lisboa -- UL, Lisboa, Portugal}

\author{M.~Aglietta}
\affiliation{Osservatorio Astrofisico di Torino (INAF), Torino, Italy}
\affiliation{INFN, Sezione di Torino, Torino, Italy}

\author{J.M.~Albury}
\affiliation{University of Adelaide, Adelaide, S.A., Australia}

\author{I.~Allekotte}
\affiliation{Centro At\'omico Bariloche and Instituto Balseiro (CNEA-UNCuyo-CONICET), San Carlos de Bariloche, Argentina}

\author{K.~Almeida Cheminant}
\affiliation{Institute of Nuclear Physics PAN, Krakow, Poland}

\author{A.~Almela}
\affiliation{Instituto de Tecnolog\'\i{}as en Detecci\'on y Astropart\'\i{}culas (CNEA, CONICET, UNSAM), Buenos Aires, Argentina}
\affiliation{Universidad Tecnol\'ogica Nacional -- Facultad Regional Buenos Aires, Buenos Aires, Argentina}

\author{R.~Aloisio}
\affiliation{Gran Sasso Science Institute, L'Aquila, Italy}
\affiliation{INFN Laboratori Nazionali del Gran Sasso, Assergi (L'Aquila), Italy}

\author{J.~Alvarez-Mu\~niz}
\affiliation{Instituto Galego de F\'\i{}sica de Altas Enerx\'\i{}as (IGFAE), Universidade de Santiago de Compostela, Santiago de Compostela, Spain}

\author{R.~Alves Batista}
\affiliation{IMAPP, Radboud University Nijmegen, Nijmegen, The Netherlands}

\author{J.~Ammerman Yebra}
\affiliation{Instituto Galego de F\'\i{}sica de Altas Enerx\'\i{}as (IGFAE), Universidade de Santiago de Compostela, Santiago de Compostela, Spain}

% ...

\collaboration{The Pierre Auger Collaboration}
\email{spokespersons@auger.org}
\homepage{http://www.auger.org}
\noaffiliation


\date{\today}

\begin{abstract}
 
\noindent Using the data of the Pierre Auger Observatory, we report on a search for signatures that would be suggestive of super-heavy particles decaying in the Galactic halo. From the lack of signal, we present upper limits for different energy thresholds above ${\gtrsim}10^8$\,GeV on the secondary by-product fluxes expected from the decay of the particles. Assuming that the energy density of these super-heavy particles matches that of dark matter observed today, we translate the upper bounds on the particle fluxes into tight constraints on the couplings governing the decay process as a function of the particle mass. Instantons, which are non-perturbative solutions to Yang-Mills equations, can give rise to decay channels otherwise forbidden and transform stable particles into meta-stable ones. Assuming such instanton-induced decay processes, we derive a bound on the reduced coupling constant of gauge interactions in the dark sector: $\alpha_X  \alt 0.09$, for $10^{9} \alt M_X/\text{GeV} < 10^{19}$. Conversely, we obtain that, for instance, a reduced coupling constant $\alpha_X = 0.09$ excludes masses $M_X \gtrsim 3\times 10^{13}~$GeV. In the context of dark matter production from gravitational interactions alone during the reheating epoch, we derive constraints on the parameter space that involves, in addition to $M_X$ and $\alpha_X$, the Hubble rate at the end of inflation, the reheating efficiency, and the non-minimal coupling of the Higgs with curvature.  

\end{abstract}

\maketitle

\section{Introduction}
\label{sec:intro}

The aim of this study is to search for signatures of Planckian-interacting massive particles in the data of the Pierre Auger Observatory and to derive constraints on the particle physics and cosmological parameters governing the viability of the Planckian scenario of dark matter (DM).  Ultra-high energy cosmic rays (UHECRs), those cosmic rays with energies above ${\simeq}10^8$\,GeV, are charged particles accelerated by electromagnetic fields in special astrophysical environments. Still, the search for subdominant fluxes of particles that could reveal either some new mechanism of particle acceleration or new physics is continuously gaining sensitivity with the increased exposure of the Pierre Auger Observatory~\cite{PierreAuger:2015eyc}. Should one detect UHECRs, and in particular photons, clustered preferentially in the direction of the Galactic Center, then this could provide compelling evidence of the presence of super-heavy relics produced in the early universe and decaying today~\cite{Bhattacharjee:1999mup,Anchordoqui:2018qom}. Such super-heavy particles have been proposed to form the DM~\cite{Ellis:1990nb,PhysRevLett.79.4302,Chung:1998zb,Kuzmin:1998kk,Chung:1999ve,Kolb:2007vd,Fedderke:2014ura,Garny:2015sjg,Ellis:2015jpg,Kolb:2017jvz,Dudas:2017rpa,Kaneta:2019zgw,Mambrini:2021zpp}.

The nature of DM remains elusive. The leading benchmark relies on assuming the existence of weakly-interacting massive particles (WIMPs) that were in equilibrium in the thermal bath of the early universe before dropping out of equilibrium when the temperature became lower than their mass~\cite{Hut:1977zn,Lee:1977ua,Vysotsky:1977pe}. To explain the relic abundance of DM observed today, the mass of these particles should lie in the range $10^2$--$10^4$~GeV, which is consistent with the expectations from the technical naturalness to have new physics at the TeV scale~\cite{tHooft:1979rat}. However, WIMPs have escaped any detection so far~\cite{MarrodanUndagoitia:2015veg,Rappoccio:2018qxp,Gaskins:2016cha}. All in all, the various null results give increasingly strong constraints for the WIMPs to match the relic density. Although the exploration of the complete WIMP parameter space remains of great importance, the current lack of signal provides a motivation to consider alternative models of DM. 

There are good motives for considering super-heavy DM (SHDM) particles rather than WIMPs. New physics could manifest only at a very high energy scale, such as the GUT scale ($M_\text{GUT}$) or even the Planck scale ($M_\text{Pl}$). Such a possibility has emerged from the estimation of the instability scale $\LambdaI$ of the Standard Model (SM) that characterizes the scale at which the SM Higgs potential develops an instability at large field values. For the current values of the Higgs and top masses and the strong coupling constant, the range of $\LambdaI$ turns out to be high, namely $10^{10}$ to $10^{12}$\,GeV~\cite{Buttazzo:2013uya,Alekhin:2012py,Bednyakov:2015sca}. While the change of sign of the Higgs quartic coupling $\lambda$ at that scale could trigger a vacuum instability due to the Higgs potential suddenly becoming unbounded from below, the running of $\lambda$ for energies above $\LambdaI$ turns out to be slow~\cite{Buttazzo:2013uya}. This peculiar behaviour leaves the possibility of extrapolating the SM to even higher energies than $\LambdaI$, up to $\MPl$, with no need to introduce new physics to stabilize the SM. In this case, the mass spectrum of the dark sector could reflect the high energy scale of the new physics. 

Various mechanisms taking place at the end of the inflationary era in Big Bang cosmology are capable of producing SHDM particles. Inflation could be driven by the presence of a scalar field, the inflaton, which slowly rolled down its potential during the inflationary era before reaching its minimum. The inflaton field then started coherent oscillations  around its minimum potential and subsequently decayed into SM particles that reheated the universe (the reheating era) while thermalizing. The production of SHDM could have occurred in the same manner on the condition that the inflaton experienced a steep potential right after the period of slow-rolling motion so as to generate large-amplitude oscillations (see, e.g.,~\cite{Chung:1998rq}). The coupling between the inflaton and the particles is however required to be fine-tuned to a very small value to avoid overshooting the DM content. Alternatively, SHDM could also be produced during the coherent oscillations of the inflaton prior to its decay, due to the ``non-adiabatic'' expansion of the background space-time acting on the vacuum quantum fluctuations~\cite{Kuzmin:1999zk,Chung:1999ve}. Particles with masses of the order of the inflaton mass can result from this gravitational production mechanism. Constraints on such scenarios have already been placed using cosmic-ray data at ultra-high energies~\cite{Aloisio:2015lva}, and will be updated and complemented in a forthcoming publication. In this article and the accompanying Letter~\cite{PierreAuger:2022wzk}, we instead consider particles with masses anywhere between ${\simeq}10^8$\,GeV and $M_\text{Pl}$. These can have been produced after the period of inflation has ended by annihilation of SM particles through the exchange of a graviton~\cite{Garny:2015sjg}, or by annihilation of inflaton particles through the same exchange of a graviton~\cite{Mambrini:2021zpp}. In this context, the only interaction between SM and dark sectors is  gravitational. For this reason, these SHDM particles have been dubbed as Planckian-interacting massive particles (PIDM), and we shall use this term hereafter when we need to be specific to this minimal coupling for SHDM particles -- keeping the term SHDM for setups with additional feeble couplings. The absence of DM-SM couplings is consistent with the large panoply of observational evidence for the existence of DM based on gravitational effects alone. Once SM and inflaton particles have populated the dark sector prior to the radiation-dominated era, the abundance of PIDM particles set by the freeze-in mechanism~\cite{McDonald:2001vt,Hall:2009bx,Bernal:2017kxu} can evolve to match the relic abundance of DM inferred today for viable parameters governing the thermal history and geometry of the universe~\cite{Garny:2015sjg}. 
 
The absence of direct coupling between PIDM and the SM (apart from gravitational) leaves only a few possible observational signatures. The large values of the Hubble expansion rate at the end of inflation $\Hinf$ needed to match the relic abundance $\Omega_\text{CDM}h$ imply tensor modes in the cosmological microwave background anisotropies that could be observed in the future~\cite{Garny:2015sjg}. On the other hand, even if the absence of direct interactions guarantees the stability of the particles in the perturbative domain, PIDM protected from decay by a symmetry can eventually disintegrate due to non-perturbative effects in non-abelian gauge theories and produce UHECRs such as (anti-)protons/neutrons, photons and (anti-)neutrinos. The aim of this study is to search for such signatures in the data from the Pierre Auger Observatory and to derive constraints on the various particle-physics and cosmological parameters governing the viability of the PIDM scenario for DM. 

The paper is organized as follows. In section~\ref{sec:auger}, we derive upper limits on the flux of secondary by-products expected from the decay of the particles. We show in particular that the most stringent limits are provided by the absence of UHE photons. By relating, in the framework of instanton-induced decay, the lifetime of the particles to the coupling constant $\alpha_X$ of a hidden sector pertaining to PIDM, the limits obtained in section~\ref{sec:auger} are shown in section~\ref{sec:gauge_constraints} to be sufficient to provide upper bounds on $\alpha_X$ as a function of $M_X$. Here $\alpha_X$ is the gauge coupling constant of a hidden non-abelian symmetry possibly unified with SM interactions at a high scale. In section~\ref{sec:eps-aX_constraints}, we use the results obtained in~\cite{Garny:2015sjg,Mambrini:2021zpp} for the PIDM scenario to relate the reheating temperature $\Trh$ (the temperature at the end of the reheating era), the Hubble expansion rate $\Hinf$ and the mass of the particles $M_X$ to the relic abundance $\Omega_\text{CDM}h=(0.1199\pm 0.0022)$~\cite{Planck:2015mrs}, with $h$ being the dimensionless Hubble constant~\cite{Planck:2015mrs}. The relationship obtained is then used to delineate viable regions to these quantities and $\alpha_X$. In parallel, it is important to assess the possible impacts of inflationary cosmologies on the astronomically-long lifetime of the vacuum of the SM~\cite{Buttazzo:2013uya,Andreassen:2017rzq}. Large fluctuations of free fields generated by the dynamics on a curved background, because of the presence of a non-minimal coupling $\xi$ between the Higgs field and the curvature of space-time, might indeed challenge this lifetime. Requiring the electroweak vacuum not to decay yields constraints between the non-minimal coupling and the Hubble rate at the end of inflation~\cite{Markkanen:2018bfx}, which are propagated in the plane $(\xi,\alpha_X)$ in section~\ref{sec:xi-aX_constraints}. Finally, the results are summarized in section~\ref{sec:conclusion}.

\section{Searches for SHDM/PIDM signatures at the Pierre Auger Observatory}
\label{sec:auger}

Regardless of the underlying model of particle physics that regulates the decay process of the SHDM particles, pairs of quarks and anti-quarks of any flavor are expected as by-products of disintegration. They give rise to a direct production of fluxes of UHE photons and neutrinos as well as to a cascade of partons that then produce a cascade of hadrons, among which are nucleons and pions, which themselves decay and generate copious fluxes of UHE photons and neutrinos. All these secondaries can be scrutinized in UHECR data. 

\subsection{Prediction of the fluxes of secondaries}
\label{sec:flux_secondaries} 

Secondaries are expected to be emitted isotropically, in proportion to the DM density accumulated in galaxy halos. For each particle $i=\{\gamma,\nu,\overline{\nu},N,\overline{N}\}$, the flux as observed on Earth is dominated by the contribution of the Milky Way halo. It can be obtained by integrating the position-dependent emission rate $q_i$ per unit volume and unit energy along the path in the direction $\nn$,
\begin{equation}
    \label{eqn:Jgal}
    J_i(E,\nn)=\frac{1}{4\pi}\int_0^\infty \dif s~q_i(E,\mathbf{x}_\odot+\mathbf{x}_i(s;\nn)).
\end{equation}
Here, $\mathbf{x}_\odot$ is the position of the Solar system in the Galaxy, $s$ is the distance from $\mathbf{x}_\odot$ to the emission point, and $\nn\equiv\nn(\ell,b)$ is a unit vector on the sphere pointing to the longitude $\ell$ and latitude $b$, in Galactic coordinates. The $4\pi$ normalisation factor accounts for the isotropy of the decay processes.

\begin{figure}
\centering
\includegraphics[width=\columnwidth]{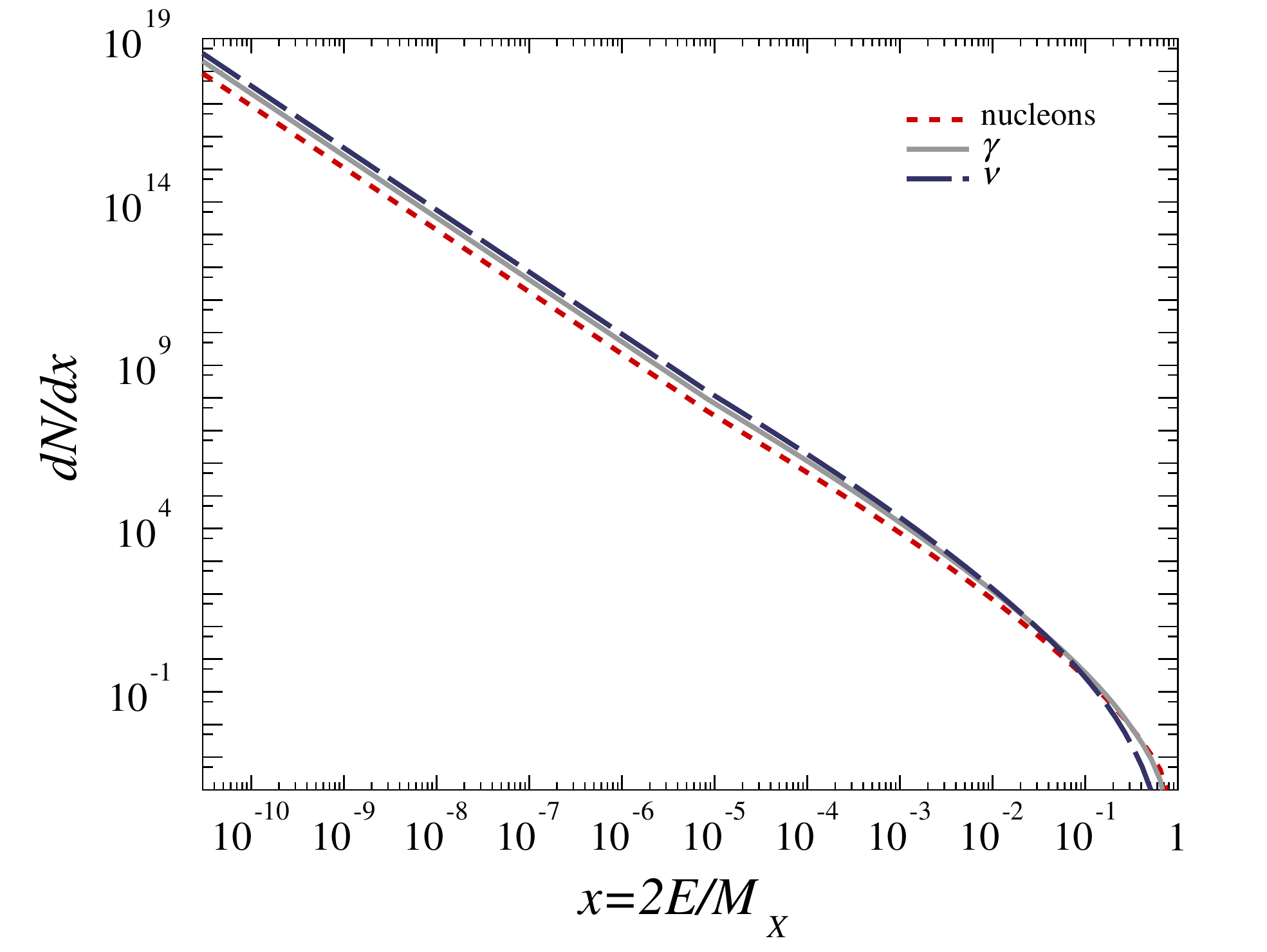}
\caption{Energy spectra of decay by-products of an SHDM particle ($M_X=M_\text{Pl}$ here) in the $q\bar q$ channel, based on the hadronization process described in~\cite{Aloisio:2003xj}.}
\label{fig:dNdx}
\end{figure}

The emission rate is shaped by the DM density $n_\text{DM}$, more conveniently expressed in terms of energy density $\rho_\text{DM}=M_X\,n_\text{DM}$, and by the differential decay width into the particle species $i$ as
\begin{equation}
    \label{eqn:q_i}
    q_i(E,\mathbf{x})=\frac{\rho_\text{DM}(\mathbf{x})}{M_X}\frac{\dif\Gamma_i(E;M_X)}{\dif E}.
\end{equation}
The ingredients are thus well separated in terms of astrophysical and particle-physics inputs. There are uncertainties in the determination of the profile $\rho_\text{DM}$. We use here the traditional NFW profile as a reference~\cite{Navarro:1995iw},
\begin{equation}
    \label{eqn:NFW}
    \rho_\text{DM}(R)=\frac{\rho_s}{(R/R_s)(1+R/R_s)^2},
\end{equation}
where $R$ is the distance to the Galactic center, $R_s=24$\,kpc, and $\rho_s$ is fixed by the DM density in the solar neighborhood, namely $\rho_\odot=0.3$\,GeV\,cm$^{-3}$. There are uncertainties in the determination of this profile. We will use other profiles such as those from Einasto~\cite{Einasto:1965czb}, Burkert~\cite{Burkert:1995yz} or Moore~\cite{Moore:1999nt} as sources of systematics. The other ingredient shaping the emission rate is the particle-physics factor that regulates the fluxes of secondary UHECRs from the decay of the super-heavy particles. In most of SHDM models, the decay is assumed to occur initially in the parton/anti-parton channel (refereed to as $q\overline{q}$ channel). The factor is then the (inclusive) differential decay width into secondary $i$ that accounts for the parton cascade and hadronization process. For a particle with mass $M_X$ decaying into partons $a$ that hadronize into particles of type $h$, the differential width $\dif \Gamma_i/\dif E$ relies primarily on the hadron energy spectrum, which can be written as~\cite{Ellis:1996mzs}
\begin{equation}
    \label{eqn:dGammadx}
    \frac{\dif N_h(x,M^2,M_X^2)}{\dif x}=\sum_a \int_x^1\frac{\dif z}{z}\frac{1}{\Gamma_a}\frac{\dif \Gamma_a(y,M_X^2)}{\dif y}\Bigg\vert_{y=x/z} D_a^h(z,M^2).
\end{equation}
Here, $x=2E_h/M_X$, $z=E_h/E_a$ and $y=x/z$ are the various fractions of available maximum momentum and primary parton momentum carried by the hadron under scrutiny. To lowest order for a two-body decay, the decay width of the particle into parton $a$, $\dif\Gamma_a/\dif y$, is proportional to $\delta(1-y)$, so that $\dif N_h/\dif x$ is then proportional to $\sum_a D_a^h(x,M^2)$, the constant of proportionality being the inverse of the number of quark flavors $n_\mathrm{F}$~\cite{Basu:2004pv}. The $D_a^h(z,M^2)$ functions are the fragmentation functions for hadrons of type $h$ from partons $a$, with $M^2$ the factorisation scale chosen to be $M^2\simeq M_X^2$. These functions are evolved, starting from measurements at the electroweak scale up to the energy scale fixed by $M_X$, using the DGLAP equation to account for the splitting function that describes the emission of parton $k$ by parton $j$. The energy spectra of photons, neutrinos and nucleons, $\dif N_i/\dif x$ with $i=\{\gamma,\nu,N\}$, then follow from the subsequent decay of unstable hadrons. Among the various computational schemes~\cite{Sarkar:2001se,Barbot:2002gt,Aloisio:2003xj,Kachelriess:2018rty,Alcantara:2019sco}, there is a general agreement for these spectra to be of the form $E^{-1.9}$. We use the scheme of Ref.~\cite{Aloisio:2003xj} in this study, which is illustrated for the quark/anti-quark channel in Fig.~\ref{fig:dNdx} in terms of $\dif N_i/\dif x$. Note that to study decays into $p$ quarks/anti-quarks pairs ($p>1)$, the phase space factor entering into Eq.~\eqref{eqn:dGammadx} through the width $\dif\Gamma_a/\dif y$ then scales as $(2p-1)(2p-2)z(1-z)^{2p-3}$~\cite{Sarkar:2001se}. 

All in all, this allows us to express $q_i$ as
\begin{equation}
    \label{eqnb:q_i}
    q_i(E,\mathbf{x})=\frac{\rho_\text{DM}(\mathbf{x})}{M_X\tau_X} \frac{\dif N_i(E;M_X)}{\dif E},
\end{equation}
with $\tau_X=\Gamma^{-1}_X$ the lifetime of the $X$ particles. The salient features of the flux from the decay by-products of super-heavy particles are thus the presence of 2-to-3 (3-to-4) times more photons (neutrinos) than nucleons on the one hand, and its peculiar directional dependency. 

\subsection{Search for secondaries from the decay of SHDM in data of the Observatory}
\label{sec:search_in_auger}

\begin{figure}
\includegraphics[width=\columnwidth]{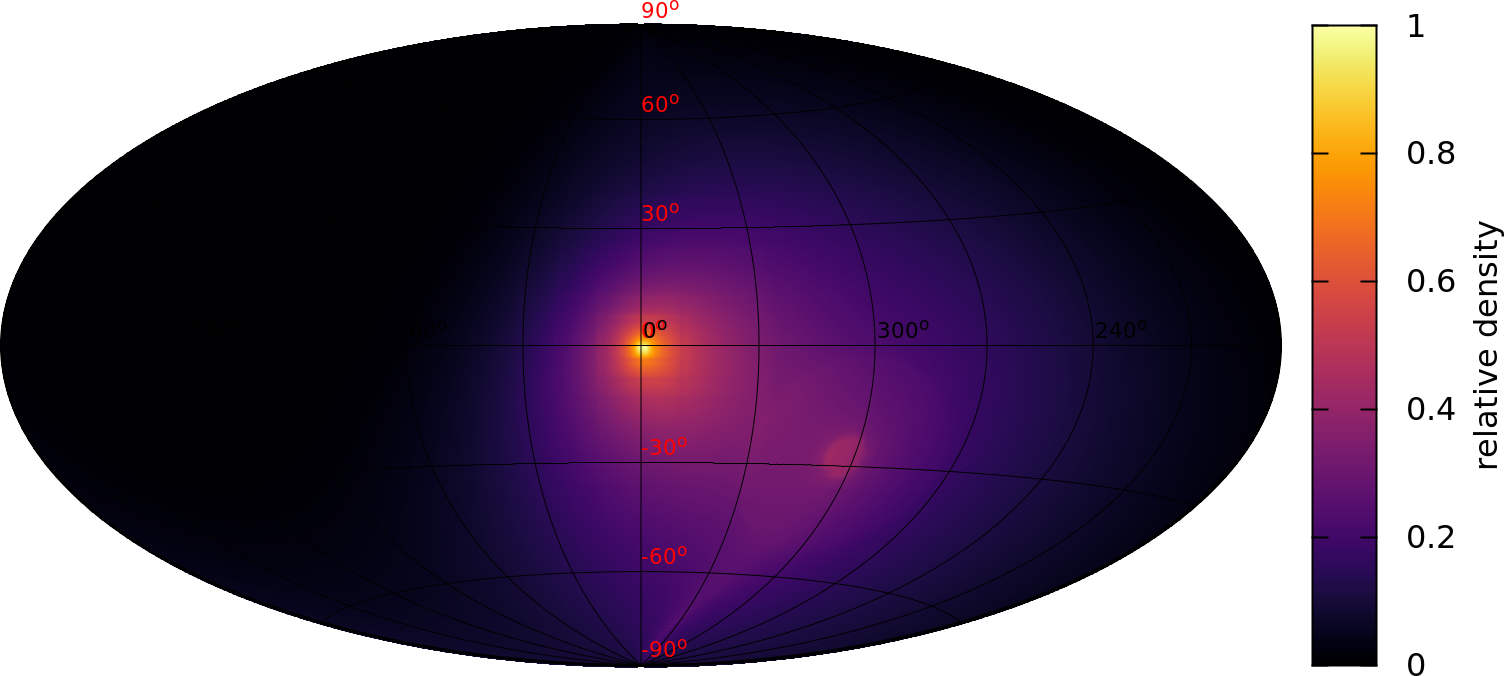}
\caption{Signal term of the directional density, $\delta\mu(\nn,E=32\,\text{EeV})$, as expected to be observed at the Pierre Auger Observatory in galactic coordinates.}
\label{fig:anismap}
\end{figure}

The features described above can give rise to observational signatures that can be captured at the Pierre Auger Observatory, located in the province of Mendoza (Argentina) and covering 3000\,km$^2$~\cite{PierreAuger:2015eyc}. UHECRs can only be studied through the detection of the showers of particles they create in the atmosphere. As the cascade develops, nitrogen and oxygen molecules get excited by the many ionizing electrons created along the shower track. The ultraviolet fluorescence caused by the subsequent de-excitation of the molecules can be detected by telescope stations, made up of arrays of several hundreds of photomultiplier tubes that, thanks to a set of  mirrors, each monitor a small portion of the sky. The isotropic emission enables observing the cascades side-on up to 30 or 40~km away on moonless nights and thus the reconstruction of the longitudinal profile of the showers. This reconstruction allows the inference of both the energy of the showers in a calorimetric way, without recourse to external information to calibrate the energy estimator, and the slant depth of maximum of shower development, ($X_\text{max}$), which is a proxy, the best available to now, of the primary mass of the particles. Complementing the fluorescence detectors, particle detectors deployed on the ground can be operated with a quasi-permanent duty cycle and thus provide a harvest of data. The subset of events detected simultaneously by the fluorescence and the surface detectors is used to develop a calibration curve such that an energy estimate can be assigned to each event~\cite{PierreAuger:2020qqz,PierreAuger:2021hun,TelescopeArray:2012qqu}. Such a hybrid-detection approach is advantageous for providing a calorimetric estimate of the energy for events recorded during periods when the telescopes cannot be operated, thus avoiding assumptions about the primary mass and the hadronic processes that control the shower development to infer the energies.

The Pierre Auger Observatory is such a hybrid system. The array of particle detectors is made of 1600 water-Cherenkov detectors deployed on a 1500\,m triangular grid. The array is overlooked from four stations, each containing six telescopes used to detect the emitted fluorescence light. The energy resolution achieved is 10\% above $10^{10}$\,GeV~\cite{PierreAuger:2020qqz}. The integrated exposure of the Observatory over the last 17 years, 122\,000~km$^2$\,sr\,yr, has enabled us to measure the arrival directions, within 1$^\circ$~\cite{PierreAuger:2020yab}, of more than 2\,600 UHECRs above $3.2{\times}10^{10}$\,GeV. This data set, the largest available at such energies, is used to search for a component of UHECRs following the arrival direction pattern predicted by Equation~(\ref{eqn:Jgal}). Previous related searches have been conducted using much more modest data sets~\cite{Berezinsky:1998rp,Benson:1999ie,Medina-Tanco:1999ktg,Evans:2001rv,Aloisio:2006yi,Siffert:2007zza}. The high energy thresholds considered here, namely from $10^{10.5}$\,GeV to $10^{10.9}$\,GeV, allow us to minimize the uncertainties inherent in the modelling of the Galactic magnetic field bending the (anti-)proton trajectories. A thorough exploration of the entire energy range accessible to the Observatory is left for a future study. 

\begin{figure}
\centering
\includegraphics[width=\columnwidth]{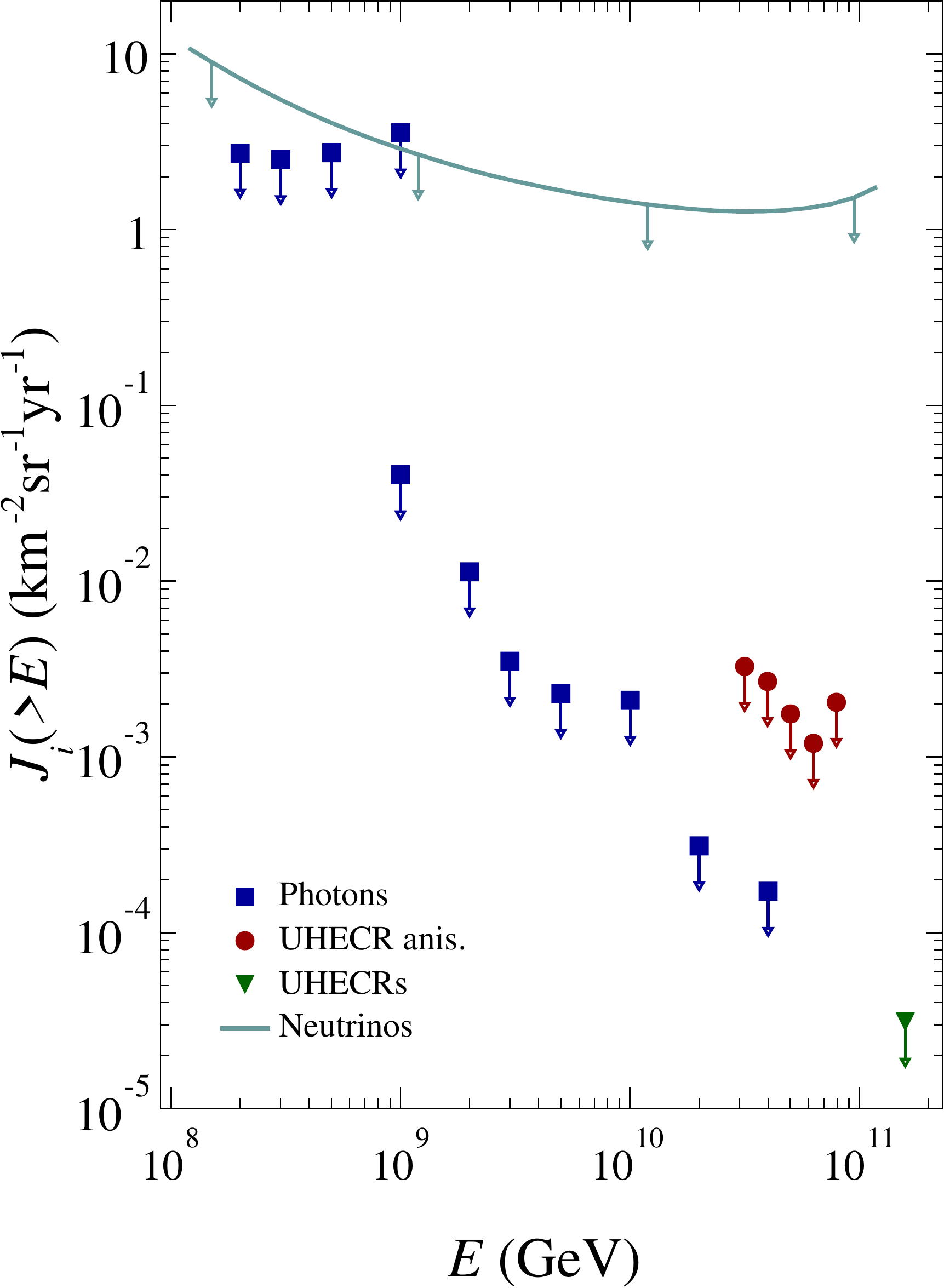}
\caption{Upper limits on secondaries produced from the decay of SHDM particles.}
\label{fig:upplim-sec}
\end{figure}

To search for a sub-dominant directional dependency suggestive of a DM signal, the set of observed arrival directions is required to match in the best possible way a directional density $\mu(\nn,E)\equiv\mu(\nn,{>}E)$ that consists of the sum of a background density and a signal density built from Eq.~\eqref{eqn:Jgal}. The balance between the two contributions is left free and denoted as $\zeta$. As the dependencies with energy of the background and of the signal terms are different, the resolution effects (in energy) are expected to distort the balance parameter. A forward-folding of the detector effects is thus carried out to build $\mu(\nn,E;\zeta)$. Under these conditions, the isotropic background density above an energy threshold $E$, $\mu_\text{bkg}(\nn,E;\zeta)$, is modelled as
\begin{widetext}
\begin{equation}
\label{eqn:mu0}
\mu_\text{bkg}(\nn,E;\zeta)=\omega(\nn)\int_{{>}E}\dif E'\int\dif E_0~ J_{\mathrm{bkg}}(E_0;\zeta)~\kappa_\text{bkg}(E',E_0),
\end{equation}
where $\omega(\nn)$ is the directional exposure~\cite{Sommers:2000us}, $J_\text{bkg}(E_0;\zeta)$ is the energy spectrum of the background built such that the total energy spectrum $J(E)$ reported in Ref.~\cite{PierreAuger:2020qqz} is the sum of the background and the signal contributions,
\begin{equation}
\label{eqn:Jbkg}
J_\text{bkg}(E_0;\zeta)=J(E)-\frac{\zeta}{4\pi}\int\dif\nn~\sum_iJ_i(E,\nn),
\end{equation}
and $\kappa_\text{bkg}(E',E_0)$ is the response function of the detector. In the energy range of interest, the latter reduces to a pure resolution function~\cite{PierreAuger:2020qqz}. The signal term, on the other hand, is given by
\begin{equation}
\delta\mu(\nn,E;\zeta)=\zeta\omega(\nn)\int_{{>}E}\dif E'\int\dif E_0~\sum_iJ_i(E_0,\nn)\kappa_i(E',E_0).
\end{equation}
\end{widetext}
Both the response function and the ``lookback position'' of the particles in the Galaxy detected in the direction $\nn$, $\mathbf{x}_i(s;\nn)$, depend on the nature of the particles:
\begin{itemize}
\item photons: a resolution function $\kappa_\gamma$ accounts for a bias (factor 2 at 30\,EeV decreasing smoothly to 1 at 100\,EeV)~\cite{PierreAuger:2022sdphotons}, while the lookback position is via straight-line motion, $\mathbf{x}_\text{n}(s)=s\nn$. 
\item (anti-)neutrons: the resolution function is approximated by that of the background,  $\kappa_\text{n}=\kappa_\text{bkg}$, while the lookback position is via straight-line motion, $\mathbf{x}_\text{n}(s)=s\nn$. The attenuation is neglected given the large decay-length value in the energy range scrutinized.
\item (anti-)protons: the resolution function is approximated by that of the background, $\kappa_\text{p}=\kappa_\text{bkg}$, while the lookback position is using the well-established method that consists of retro-propagating protons and anti-protons from the Earth, counting the time spent in $\rho_\text{DM}$ before exiting the Galaxy~\cite{1968JPhA....1..694T}. The magnetic field model contains the so-called JF12 regular component~\cite{Jansson:2012rt} and a turbulent one, the amplitude of which is fixed to equal the envelope of the regular field. 
\item (anti-)neutrinos: they are not accounted for in this anisotropy-search analysis, given the absence of a contribution to the observed number of events.  
\end{itemize}
The resulting density $\delta\mu(\nn,E)$ is shown in Fig.~\ref{fig:anismap} for $E=32$\,EeV. The final density fitted to the data through a likelihood function $L(\zeta)=\prod_\text{events}\mu(\nn_i,E;\zeta)$ is normalised to 1 when integrated over arrival directions,
\begin{equation}
\mu(\nn,E;\zeta)=\frac{\mu_\text{bkg}(\nn,E;\zeta)+\delta\mu(\nn,E;\zeta)}{\int\dif\nn\,\mu_0(\nn,E;\zeta)+\int\dif\nn\,\delta\mu(\nn,E;\zeta)}.
\end{equation}
The analysis is performed for energy thresholds spaced by $\Delta\lg{E}=0.1$. The largest deviation from the no-signal hypothesis is insignificant (within $2\sigma$) for $\lg{(E/\text{GeV})}=10.7$. Upper limits at $90\%$ C.L.\ on the all-sky-averaged  $J_\text{DM}(E)\equiv\sum_i J_i(E)$ flux are then obtained by solving with Monte Carlo simulations the equation $\int_{{\geq}\mathcal{L}_\text{data}}\dif\mathcal{L}~p(\mathcal{L}(\zeta_{90}))=0.90$ and are reported as the red filled circles in Fig.~\ref{fig:upplim-sec}.

Apart from the anisotropies present in the arrival directions, another signature in favor of the decay of SHDM particles would be the presence of UHE photons in the data of the Observatory. The identification of photon primaries relies on the ability to distinguish the showers generated by photons from those initiated by the overwhelming background of nuclei. Since the radiation length in the atmosphere is more than two orders of magnitude smaller than the mean free path for photo-nuclear interactions, the transfer of energy to the hadron/muon channel in photon showers is reduced with respect to the bulk of hadron-induced showers, resulting in a lower number of secondary muons. Additionally, as the development of photon showers is delayed by the typically small multiplicity of electromagnetic interactions, they reach $X_\text{max}$ deeper in the atmosphere with respect to showers initiated by hadrons. Both the ground signal and $X_\text{max}$ can be measured at the Observatory. Although showers are observed at a fixed slice in depth with the array of particle detectors, the longitudinal development is embedded in the signals detected. The fluorescence and particle detectors are complemented with the low-energy enhancements of the Observatory, namely three additional fluorescence telescopes with an elevated field of view, overlooking a denser array of particle detectors, in which the stations are separated by 750\,m. The combination of these instruments allows showers to be measured in the energy range above $10^{8}$\,GeV. 

Three different analyses, differing in the detector used, have been developed to cover the wide energy range probed at the Observatory and have been reported in Ref.~\cite{PierreAuger:2022uwd,Savina:2021ufs,PierreAuger:2022aty}. No photons with energies above $2{\times}10^{8}$\,GeV have been unambiguously identified so far, leading to the 95\% C.L.\ flux upper limits displayed in Fig.~\ref{fig:upplim-sec} as the filled blue squares. The limit above $10^{11.2}$\,GeV (green triangle), stemming from the non-detection so far of any UHECR~\cite{PierreAuger:2020qqz}, including photons, is also constraining~\cite{Alcantara:2019sco,Anchordoqui:2021crl}. In the energy range above $2{\times}10^{10}$\,GeV, the limits on photon fluxes are observed to be much more constraining than those inferred from the absence of significant anisotropies. This is because the accumulated exposure to photons enables us to probe fluxes less than a few percent of that of UHECRs, while the current sensitivity to anisotropies does not allow for capturing an amplitude less than 10 to 15\% of the UHECR flux.

Finally, (anti-)neutrinos, another emblematic signature of SHDM particle decays, can also be identified at the Observatory. Neutrinos of all flavors can interact in the atmosphere through charged- or neutral-current interactions and induce a ``downward-going'' shower that can be detected~\cite{Capelle:1998zz}. In addition, tau neutrinos ($\nu_\tau$) can undergo charged-current interactions and produce a $\tau$ lepton in the Earth's crust that eventually decays in the atmosphere, inducing an upward-going shower
~\cite{Bertou:2001vm}. Tau neutrinos are not expected to be copiously produced at the astrophysical sources; yet approximately equal fluxes for each neutrino flavour should reach the Earth as a result of neutrino oscillations over cosmological distances~\cite{Learned:1994wg,Athar:2000yw,Anchordoqui:2013dnh}. The identification of neutrinos relies on salient zenith-dependent features of air showers. For highly-inclined cascades (zenith angle larger than 60$^\circ$), neutrino-induced showers initiated deep in the atmosphere near ground level have a significant electromagnetic component when they reach the array of particle detectors, producing signals that are spread over time. In contrast, inclined showers initiated at a shallow depth in the atmosphere by the bulk of UHECRs are dominated by muons at the ground level, inducing signals in the particle detectors that have characteristic high peaks associated with individual muons, which are spread over smaller time intervals. Thanks to the fast sampling (25\,ns) of the digital electronics of the detectors, several observables that are sensitive to the time structure of the signal can be used to discriminate between these two types of showers. 

Neutrino limits obtained at the Observatory~\cite{PierreAuger:2019ens} are also displayed in Fig.~\ref{fig:upplim-sec} as the continuous line. Except at the lowest energies, these limits are seen to be superseded by photon limits in the search for SHDM by-product decays.

\section{Constraints on gauge coupling in the dark sector}
\label{sec:gauge_constraints}

\subsection{Pertubative-decay processes}
\label{sub:pertdecay}

Some SHDM models postulate the existence of super-weak couplings between the dark and SM sectors. The lifetime $\tau_X$ of the particles is then governed by the strength of the couplings $g_{X\Theta}$ (or reduced couplings $\alpha_{X\Theta}=g_{X\Theta}^2/(4\pi)$) and by the mass dimension $n$ of the operator $\Theta$ standing for the SM fields in the effective interaction~\cite{deVega:2003hh}. Even without knowing the theory behind the decay of the DM particle, we can derive generic constraints on $\alpha_{X\Theta}$ and $n$. The effective interaction term that couples the field $X$ associated with the heavy particle to the SM fields is taken as
\begin{equation}
    \label{eqn:Lint}
    \mathcal{L}_\text{int}=\frac{g_{X\Theta}}{\Lambda^{n-4}}X\Theta,
\end{equation}
where $\Lambda$ is an energy parameter typical of the scale of the new interaction. In the absence of further details about the operator $\Theta$, the matrix element describing the decay transition is considered flat in all kinematic variables so that it behaves as $|\mathcal{M}|^2\sim 4\pi\alpha_{X\Theta}/\Lambda^{2n-4}$. On the basis of dimensional arguments, the lifetime of the particle $X$ is then given as
\begin{equation}
    \label{eqn:tauXTheta}
    \tau_{X\Theta}=\frac{V_n}{4\pi M_X\alpha_{X\Theta}}\left(\frac{\Lambda}{M_X}\right)^{2n-8},
\end{equation}
where $V_n$ is a phase space factor. As a proxy for this factor, we use the expression derived for $N-1$ particles in the final state~\cite{Kleiss:1985gy},
\begin{equation}
    \label{eqn:Vn}
    V_n=\left(\frac{2}{\pi}\right)^{n-1}\Gamma(n-1)\Gamma(n-2),
\end{equation}
with $\Gamma(x)$ the Euler gamma function.

\begin{figure}
\centering
\includegraphics[width=\columnwidth]{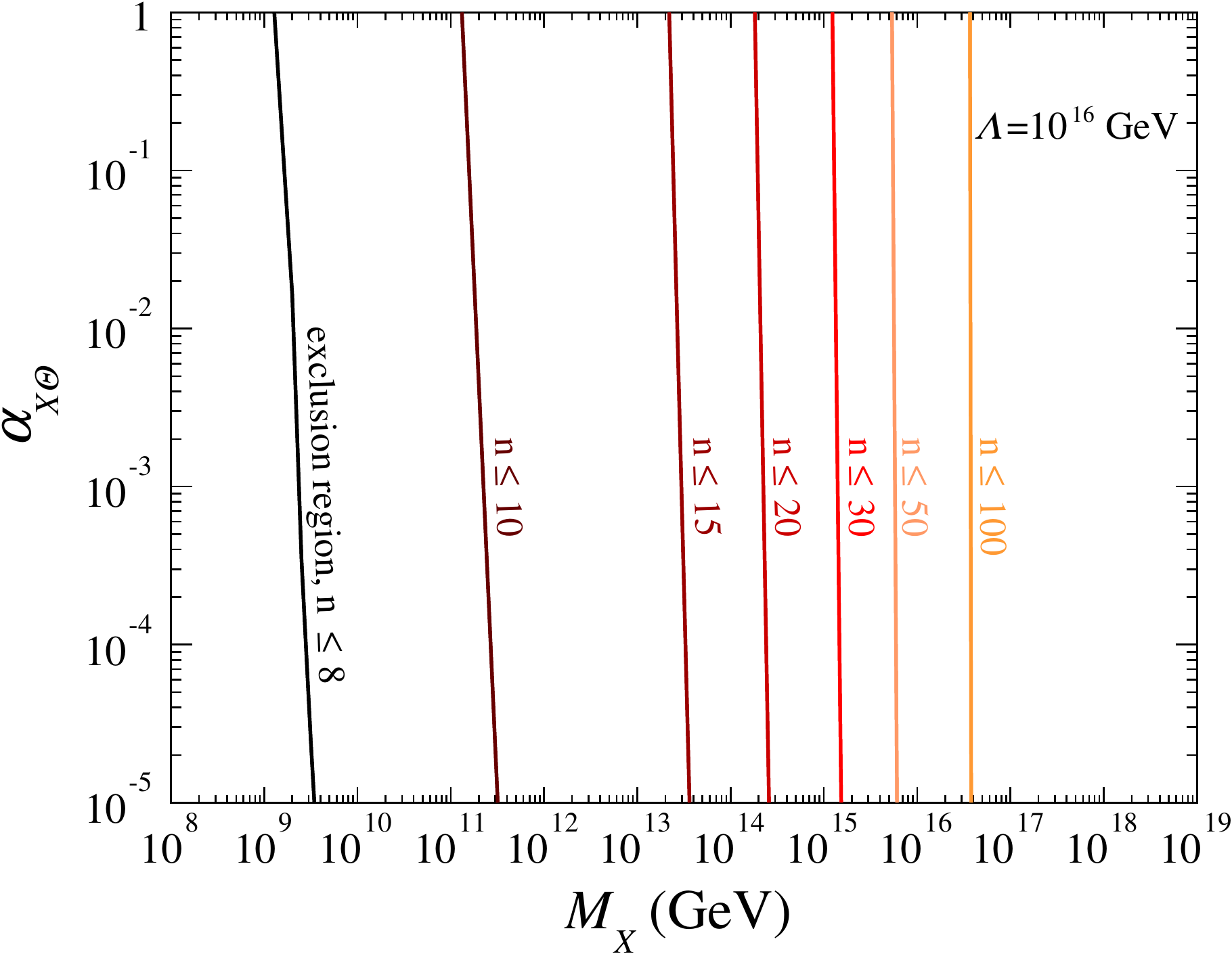}
\caption{\small{Exclusion regions in the plane $(\alpha_{X\Theta},M_X)$ for several values of mass dimension $n$ of operators responsible for the perturbative decay of the super-heavy particle, and for an energy scale of new physics $\Lambda=10^{16}$\,GeV.}} 
\label{fig:alphaX-mass-pert}
\end{figure}

Equation~(\ref{eqn:tauXTheta}) provides us with a relationship connecting the lifetime $\tau_{X\Theta}$ to the coupling constant $\alpha_{X\Theta}$ and to the dimension $n$. 

From Eq.~\eqref{eqn:tauXTheta}, it is apparent that the coupling constant $\alpha_{X\Theta}$ and the dimension $n$ have to take specific values for super-heavy particles to be stable enough~\cite{Ellis:1990nb,deVega:2003hh}. We now show that the absence of UHE photons provides powerful data to infer the viable range of values. Assuming that the relic abundance of DM is saturated by SHDM, constraints can be inferred in the plane $(\tau_{X\Theta},M_X)$ by requiring the flux calculated by averaging Equation~(\ref{eqn:Jgal}) over all directions to be less than the limits,
\begin{equation}
\label{eqn:j95}
\int_{E}^{\infty}\dif E'\langle J_{\gamma}(E',\nn)\rangle \leq J^{95\%}_{\gamma}({\geq}E),
\end{equation}
where $\langle \cdot \rangle$ stands for the average over all directions. In practice, for a specific upper limit at one energy threshold, a lower limit of the $\tau_{X\Theta}$ parameter is derived for each value of mass $M_X$. The lower limit on $\tau_{X\Theta}$ is subsequently transformed into an upper limit on $\alpha_{X\Theta}$ by means of Eq.~\eqref{eqn:tauXTheta}. This defines a curve in the plane $(\tau_{X\Theta},M_X)$. By repeating the procedure for each upper limit on $J^{95\%}_\gamma(\geq E)$, a set of curves is obtained, reflecting the sensitivity of a specific energy threshold to some range of mass. The union of the excluded regions finally provides the constraints in the plane $(\alpha_{X\Theta},M_X)$. In this manner we obtain the contour lines shown in Fig.~\ref{fig:alphaX-mass-pert} for several values of $n$ and for an emblematic choice of GUT $\Lambda$ value. The scale chosen for $\alpha_{X\Theta}$ ranges from 1 down to $10^{-5}$. It is observed that for the limits on photon fluxes to be satisfied, the mass of the super-heavy particle cannot exceed ${\gtrsim}10^9$\,GeV (${\gtrsim}10^{11}$\,GeV) for operators of dimension equal to or larger than $n=8$ ($n=10$), while larger masses require an increase in $n$. To approach the large masses while keeping operators of dimension relatively low, ``astronomically-small'' coupling constants should be at work. The same conclusions hold for other choices of $\Lambda$. All in all, for perturbative processes to be responsible for the decay of SHDM particles requires quite ``unnatural'' fine-tuning.\footnote{See, however, Ref.~\cite{Dudas:2020sbq} for a model in which SHDM couples to the neutrino sector.}

\subsection{Instanton-induced decay processes}
\label{sub:nonpertdecay}

The sufficient stability of super-heavy particles is better ensured by a new quantum conserved in the dark sector so as to protect the particles from decaying. The only interaction between the dark sector and the SM one is then gravitational, as in the PIDM instance of SHDM models.  Nevertheless, even stable particles in the perturbative domain will in general eventually decay due to non-perturbative effects in non-abelian gauge theories. Such effects, known as instantons~\cite{Belavin:1975fg,Coleman:1978ae,Vainshtein:1981wh}, provide a signal for the occurrence of quantum tunneling between distinct classes of vacua, forcing the fermion fields to evolve during the transitions and leading to the generation of particles depending on the associated anomalous symmetries~\cite{tHooft:1976rip}. Instanton-induced decay can thus make observable a dark sector of PIDM particles that would otherwise be totally hidden by the conservation of a quantum number. Following Ref.~\cite{Kuzmin:1997jua}, we assume quarks and leptons carry this quantum number and so contribute to anomaly relationships with contributions from the dark sector,\footnote{Alternatively, the particles of the dark sector could carry some SM hypercharge.} they will be secondary products in the decays of PIDM together with the lightest hidden fermion. The presence of quarks and leptons in the final state is sufficient to make usable the hadronization process described in Section~\ref{sec:auger}. The exact particle content is governed by selection rules arising from the instanton transitions that are regulated by the fermions coupled to the gauge field of the dark sector.  
As a proxy inspired from Ref.~\cite{Kuzmin:1997jua}, we assume here that a dozen of $q \bar q$ pairs are produced in the decay process and that half of the energy goes into the dark sector.

\begin{figure}
\centering
\includegraphics[width=\columnwidth]{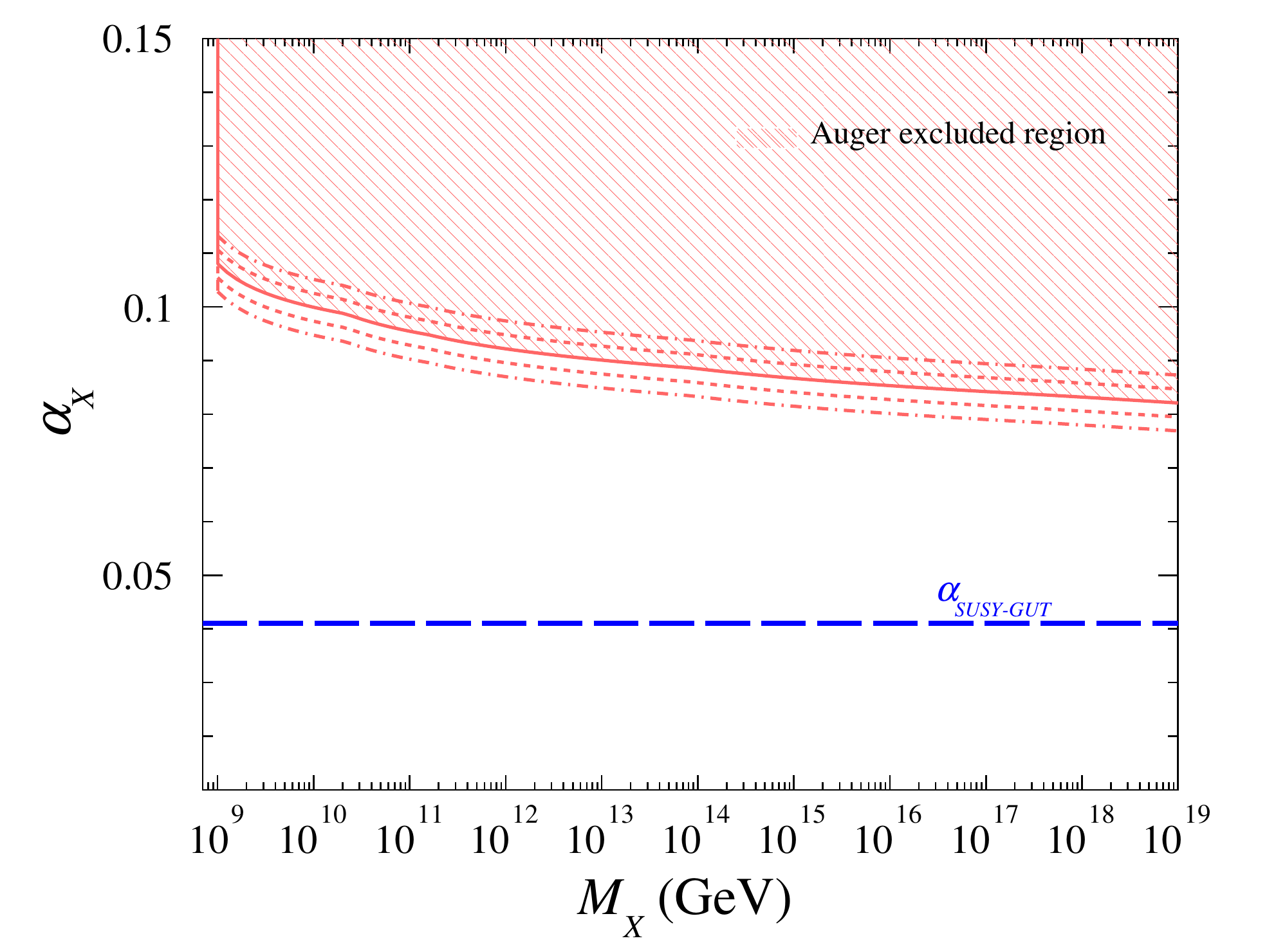}
\caption{Upper limits at 95\% C.L.\ on the coupling constant $\alpha_X$ of a hidden gauge interaction as a function of the mass $M_X$ of a dark matter particle $X$ decaying into a dozen of $q \bar q$ pairs. The dotted and dashed-dotted lines represent the systematic uncertainties stemming from the quantum fluctuations about the instanton contribution to the transition amplitude (see text). For reference, the unification of the three SM gauge couplings is shown as the horizontal blue dashed line in the framework of supersymmetric GUT~\cite{10.1093/ptep/ptaa104}.}
\label{fig:alphaX-mass}
\end{figure}

The lifetime of the decaying particle follows from the corresponding instanton-transition amplitude obtained as a semi-classical expansion of the associated path integral about the instanton solution, which provides the zeroth-order contribution that depends exponentially on $g_X^{-2}$~\cite{tHooft:1976rip}. It is the introduction of this exponential factor in the effective interaction term that suppresses to a large extent the fast decay of the particles. Considering this zeroth-order contribution only, and recasting the expression in terms of the reduced coupling constant of the hidden gauge interaction $\alpha_X$, the lifetime of the particles is given as
\begin{equation}
\label{eqn:tauX}
\tau_X\simeq M_X^{-1}\exp{\left(4\pi/\alpha_X\right)}.
\end{equation}
In this expression, we dropped, following Ref.~\cite{Kuzmin:1997jua}, the functional determinants arising from the effect of quantum fluctuations around the (classical) contribution of the instanton configurations. Those from the Yang-Mills gauge fields yield a dependency in $(4\pi\alpha_X)^{5+n_1}$ in Eq.~\eqref{eqn:tauX} with $n_1=3~(7)$ for $SU$(2) ($SU$(3)) theories for instance, a dependency that is negligible compared to the exponential one in $\alpha_X^{-1}$. Other functional determinants arise from the exact content of fields of the underlying theory. Again, the constraints inferred on $\alpha_X$ using Eq.~\eqref{eqn:tauX} are barely changed for a wide range of numerical factors given the exponential dependency in $\alpha_X^{-1}$. 

Eq.~\eqref{eqn:tauX} provides us with a relationship connecting the lifetime $\tau_X$ to the coupling constant $\alpha_X$. In the same way as in the perturbative case above, upper limits on $\alpha_X$ can be obtained. They are shown as the shaded red area in Fig.~\ref{fig:alphaX-mass}. Our results show that the coupling should be less than $\simeq 0.09$ for a wide range of masses. As already stated, numerical factors could however arise in Equation~(\ref{eqn:tauX}) depending on the underlying model for the hidden gauge sector. For example, for a theory with a hidden Higgs field responsible for mass generation in the dark sector, the factors would involve the ratio between the mass of the lightest dark state and the energy scale of new physics through the vacuum expectation value~\cite{Carone:2010ha}. Such explicit constructions of the dark sector are, however, well beyond the scope of this experimental study. Although the limits presented in Fig.~\ref{fig:alphaX-mass} are hardly destabilized due to the exponential dependence in $\alpha_X^{-1}$, we note that a shift of $\pm 0.0013k$ for factors $10^{\pm k}$ and limit ourselves to showing in dotted and dashed lines the bounds that would be obtained for $k=2$ and $k=4$, respectively. These factors are by far the dominant systematic uncertainties.

\section{Constraints on the production of PIDM particles during reheating}
\label{sec:eps-aX_constraints}

We now turn to the connection between the results presented in Fig.~\ref{fig:alphaX-mass} and the scenarios of inflationary cosmologies. In addition to the instanton-mediated decays, PIDM particles can interact gravitationally. Two recent studies~\cite{Garny:2015sjg,Mambrini:2021zpp} have shown that the gravitational interaction alone may have been sufficient to produce the right amount of DM particles at the end of the inflation era for a wide range of high masses, up to $M_\text{GUT}$. PIDM particles are naturally part of this scheme. While the observation of UHE photons could open a window to explore high-energy gauge interactions and possibly GUTs effective in the early universe, the constraints inferred on $\alpha_X$ allow us to probe the gravitational production of PIDM. We give below the main steps to derive an expression (Eq.~\eqref{eqn:abundance}) relating the present-day relic abundance of DM to the mass $M_X$ and other relevant parameters; more details can be found in Refs.~\cite{Garny:2015sjg} and~\cite{Mambrini:2021zpp}.

PIDM particles are assumed to be produced by annihilation of SM particles~\cite{Garny:2015sjg} or of inflaton particles~\cite{Mambrini:2021zpp} through the exchange of a graviton after the period of inflation has ended at time $\Hinf^{-1}$. In this context, SM particles are created by the decay of coherent oscillations of the inflaton field, $\phi$, with width $\Gamma_\phi$, which is regulated by the coupling of the inflaton to SM particles $g_\phi$ and its mass $M_\phi$ as $\Gamma_\phi=g_\phi^2M_\phi/(8\pi)$.  They subsequently scatter and thermalize until the reheating era ends at time $\Gamma_\phi^{-1}$ when the radiation-dominated era begins with temperature $\Trh$. This latter parameter, given by
\begin{equation}
    \label{eqn:Trh}
    \Trh\simeq 0.25\epsilon(\MPl\Hinf)^{1/2}
\end{equation}
with $\epsilon=(\Gamma_\phi/\Hinf)^{1/2}$ the efficiency of reheating, is obtained by assuming an instantaneous conversion of the energy density of the inflaton into radiation for a value of the cosmological scale factor $a$ such that the expansion rate $\Hinf$ equates with the decay width $\Gamma_\phi$~\cite{Kolb:1990vq}. Here, the number of degrees of freedom at reheating has been assumed to be that of the SM. For an instantaneous reheating to be effectively achieved, $\epsilon$ must approach 1, which, from the expression of $\Gamma_\phi$, requires $M_\phi$ to be of order of $\Hinf$ and $g_\phi$ not too weak. In the following, both $\Hinf$ and $\epsilon$ will be considered as free parameters to be constrained.

The dynamics of the reheating period are quite involved~\cite{Chung:1998rq,Giudice:2000ex}.\footnote{Note that we consider throughout this section, as in~\cite{Chung:1998rq,Giudice:2000ex}, an equation of state $w=0$ for the inflaton field dynamics.} As the SM particles thermalize, the plasma temperature rises rapidly to a maximum before subsequently decreasing as $T(a)\propto a^{-3/8}$,
\begin{equation}
    \label{eqn:T}
    T(a)\simeq 0.2(\epsilon\MPl\Hinf)^{1/2}\left(a^{-3/2}-a^{-4}\right)^{1/4}.
\end{equation}
The $a^{-3/8}$ scaling continues until the age of the universe is equal to $\Gamma_\phi^{-1}$, signaling the beginning of the radiation-dominated era at temperature $\Trh$. During this period, the Hubble rate $H(a)$ scales as the square root of the energy density of the inflaton, $\rho_\phi$, which itself scales as $\rho_\text{inf}(a_\text{inf}/a)^3$. Consequently, $H(a)$ evolves as $a^{-3/2}$, namely $H(a)=H_\text{inf}(a/a_\text{inf})^{-3/2}$ with $a_\text{inf}$ being the scale factor at the end of inflation. After reheating, both the temperature and the Hubble rate follow the standard evolution in a radiation-dominated era, namely $T(a)\propto \Trh a_\text{rh}/a$ and $H(a)=H_\text{inf}\epsilon^2(a/a_\text{rh})^{-2}$. The scale factor at the end of reheating is $a_\text{rh}=\epsilon^{-4/3}a_\text{inf}$, guaranteeing the continuity of $H(a)$.

\begin{figure}
\centering
\includegraphics[width=\columnwidth]{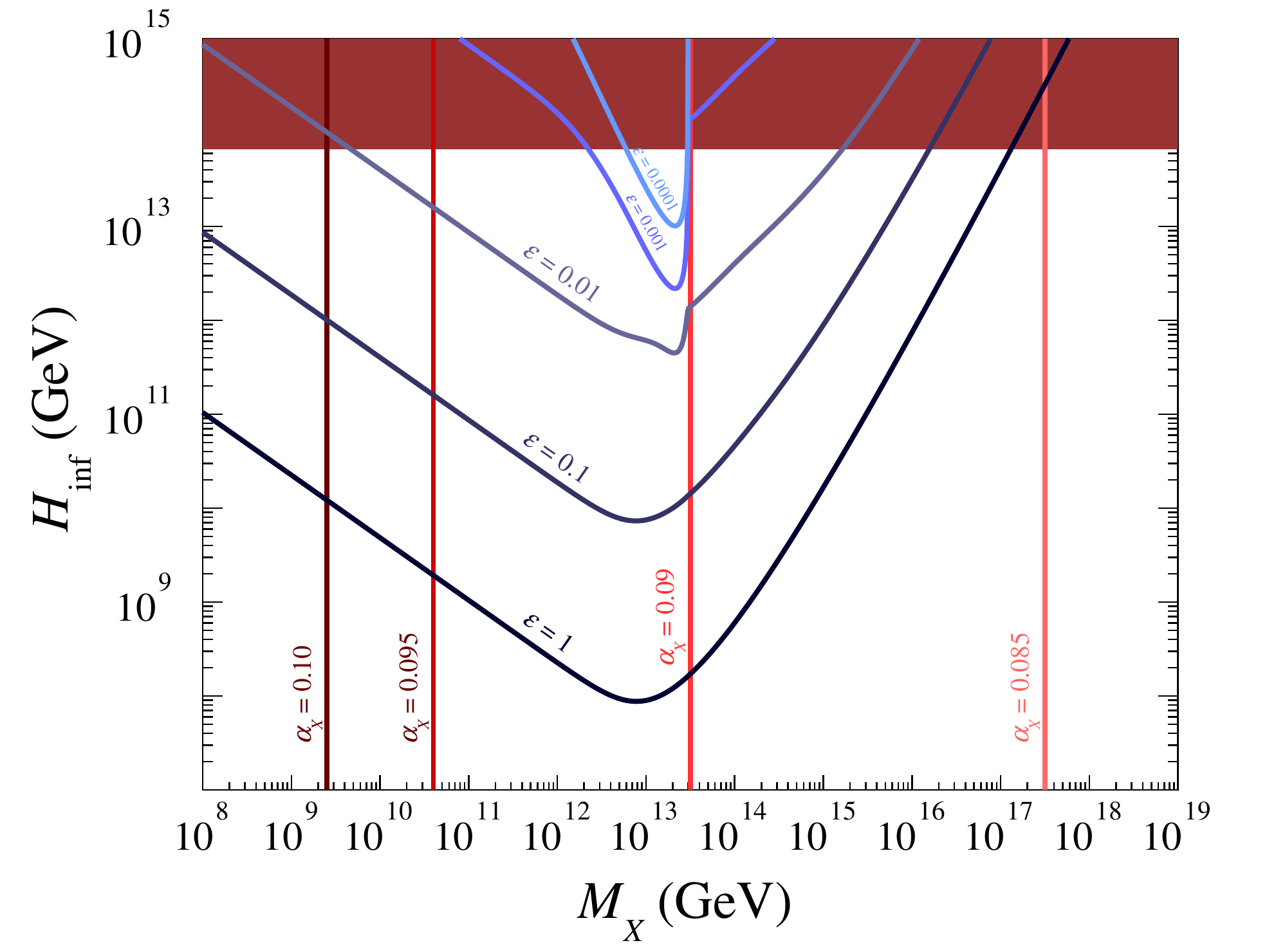}
\caption{Constraints in the $(\Hinf,M_X)$ plane. The red region is excluded by the non-observation of tensor modes in the cosmic microwave background~\cite{Garny:2015sjg,Planck:2015sxf}. The regions of viable $(\Hinf,M_X)$ values needed to set the right abundance of DM are delineated by the curves for different values of reheating efficiency $\epsilon$~\cite{Garny:2017kha} from dark blue ($\epsilon=1$) to lighter ones ($\epsilon=10^{-4}$), while values above (below) the lines lead to overabundance of (negligible quantity of) DM. Additional constraints from the non-observation of instanton-induced decay of SHDM particles allow for excluding the mass ranges in the regions to the right of the vertical lines, for the specified value of the dark-sector gauge coupling.}
\label{fig:Hinf-MX}
\end{figure}

With these reheating dynamics in hand, the relic abundance of PIDM particles can be estimated. The energy density of the universe is then in the form of unstable inflaton particles, SM radiation and stable massive particles, the time evolution of which is governed by a set of coupled Boltzmann equations~\cite{Chung:1998rq}. However, because the energy density of the massive particles is always sub-dominant, the evolution of the  inflationary and radiation energy densities largely decouple from the time evolution of the $X$-particle density $n_X$. In addition, because PIDM particles interact through gravitation only, they never come to thermal equilibrium. In this case, the collision term in the Boltzmann equation can be approximated as a source term only,
\begin{equation}
\label{eqn:nX}
\frac{\dif n_X(t)}{\dif t}+3H(t)n_X(t)\simeq \sum_i \overline{n}_i^2(t)\gamma_i.
\end{equation}
Here, the sum in the right hand side stands for the contributions from the SM and inflationary sectors. In the SM sector, $\overline{n}_i=m_X^2TK_2(M_X/T)/(2\pi^2)$~\cite{Chung:1998zb}, with $K_2(x)$ being the modified Bessel function of the second kind, and $\gamma_i=\langle\sigma v\rangle$ is the thermal-averaged cross section times velocity describing the SM+SM$\to$PIDM+PIDM reaction~\cite{Garny:2015sjg,Garny:2017kha}, which behaves as $M_X^2/\MPl^4$ for $M_X\gg T$ and as $T^2/\MPl^4$ for $M_X\ll T$. In the inflationary sector, $\overline{n}_i=\rho_\text{inf}(a_\text{inf}/a)^3/M_\phi$, with $M_\phi=3{\times}10^{13}$\,GeV in the following, and the production rate $\gamma_i$ describes the $\phi+\phi\to$PIDM+PIDM reaction~\cite{Mambrini:2021zpp}. In both SM and inflationary sectors, the production rates $\gamma_i$ for fermionic DM are considered in the following. Introducing the dimensionless abundance $Y_X=n_Xa^3/\Trh^3$ to absorb the expansion of the universe, and using $aH(a)\dif t=\dif a$ from the definition of the Hubble parameter, Eq.~\eqref{eqn:nX} becomes
\begin{equation}
\label{eqn:YX}
\frac{\dif Y_X(a)}{\dif a}\simeq\frac{a^2}{\Trh^3H(a)}\sum_i \overline{n}_i^2(a)\gamma_i,
\end{equation}
which, using the dynamics of the expansion rate during reheating described above, yields the present-day dimensionless abundance $Y_{X,0}$ assuming $Y_{X,\mathrm{inf}}=0$. The present-day relic abundance, $\Omega_\text{CDM}$, can then be related to $M_X$, $H_\text{inf}$, and $\epsilon$ through~\cite{Garny:2015sjg}
\begin{equation}
\label{eqn:abundance}
    \Omega_\text{CDM}h^2=9.2{\times}10^{24}\frac{\epsilon^4M_X}{\MPl}Y_{X,0}.
\end{equation} 

The viable $(H_\text{inf},M_X)$ parameter space is delineated by the curves corresponding to different values of $\epsilon$ in Fig.~\ref{fig:Hinf-MX}, from dark blue ($\epsilon=1$) to lighter ones ($\epsilon=10^{-4}$). As the source term in the r.h.s.\ of Eq.~\eqref{eqn:nX} raises faster with $\Hinf$ than $\Trh^3 H(a)$, $Y_X$ is a rising function of $\Hinf$, and values for $(H_\text{inf},M_X)$ above (below) the lines lead to overabundance of (negligible quantity of) DM. For high efficiencies (corresponding to short duration of the reheating era), the SM+SM$\to$PIDM+PIDM reaction allows for a wide range of $M_X$ values to fulfill Eq.~\eqref{eqn:abundance}. 
For $M_X$ to be around the GUT scale, the expansion rate $\Hinf$ (being the proxy of the energy scale of the inflation) must be sufficiently high. Arbitrarily large values of $H_\text{inf}$ are however not permitted because of the 95\% C.L.\ limits on the tensor-to-scalar ratio in the cosmic microwave background anisotropies, which, once converted into limits on the energy scale of inflation when the pivot scale exits the Hubble radius~\cite{Garny:2015sjg,Planck:2015sxf}, yield $\Hinf\leq 4.9{\times}10^{-6}\MPl$. For efficiencies below $\simeq 0.01$, the $\phi+\phi\to$PIDM+PIDM reaction allows for solutions in a narrower range of the $(H_\text{inf},M_X)$ plane, with in particular $M_X \leq M_\phi$ as a result of the kinematic suppression in the corresponding rate $\gamma_i$~\cite{Mambrini:2021zpp}. 

A clear signature of the PIDM scenario could be the detection of UHE photons produced by the instanton-induced decay of the PIDM particles -- so that no coupling between the sectors is required except gravitation. The excluded mass ranges obtained from the non-observation of instanton-induced decay of PIDM particles are regions to the right of the vertical lines for different values of dark-sector gauge coupling. While the range of $M_X$ extends from (well) below $10^8$\,GeV to ${\simeq}10^{17}$\,GeV in the case of instantaneous reheating ($\epsilon=1$) and $\alpha_X\leq 0.085$, the parameter space is observed to shrink for longer reheating duration and larger dark-sector gauge coupling. With the current sensitivity, there are no longer pairs of values $(\Hinf,M_X)$ satisfying Eq.~\eqref{eqn:abundance} for $(\epsilon\geq 0.01,\alpha_X\geq 0.10)$. 

\begin{figure}
\centering
\includegraphics[width=\columnwidth]{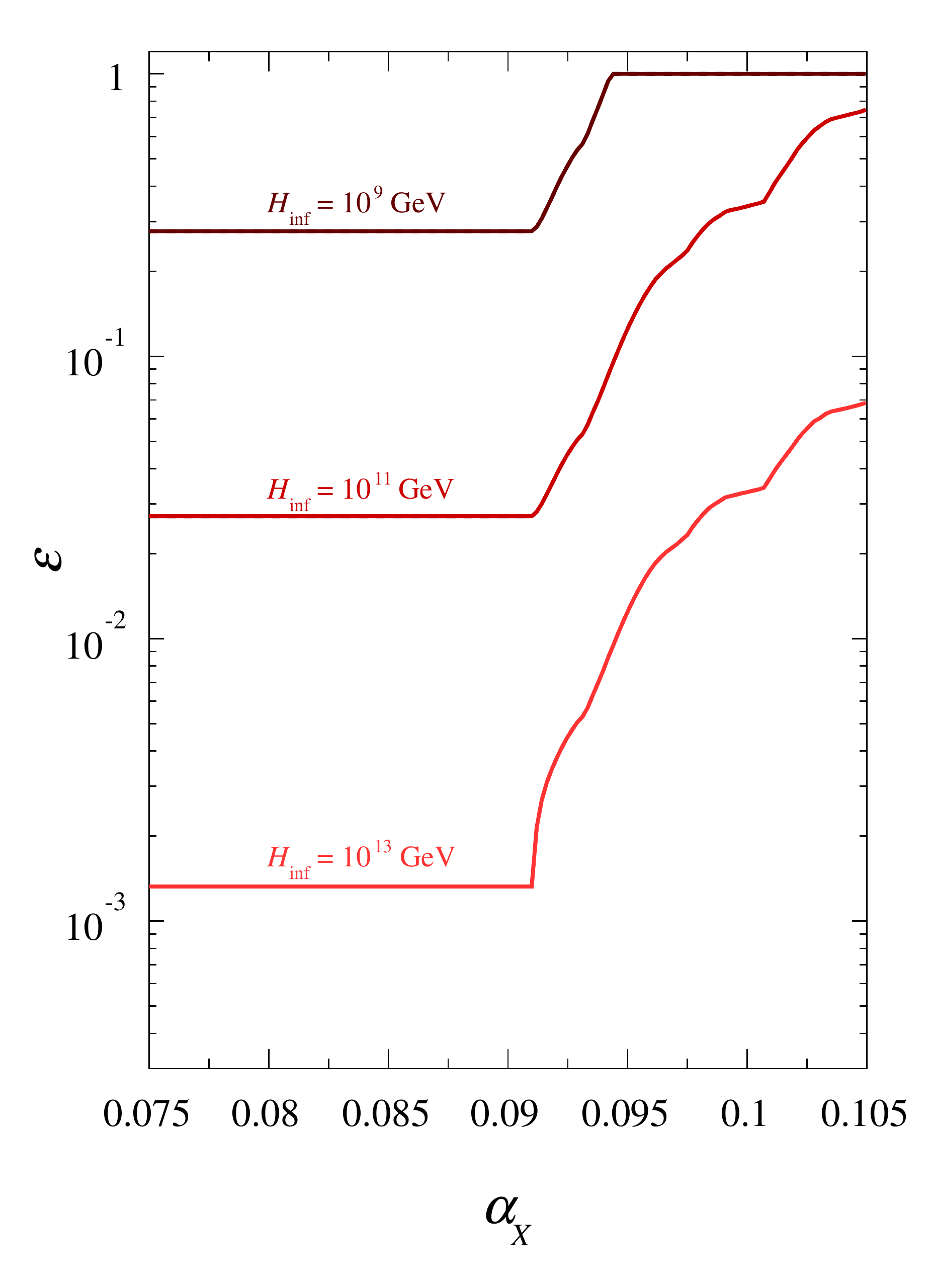}
\caption{Allowed range of $(\epsilon,\alpha_X)$ values for the scenario of Planckian-interactive massive particles as DM delineated for three examples of $\Hinf$. All regions under the lines are excluded.}
\label{fig:epsilon_aX}
\end{figure}

The allowed range of $(\epsilon,\alpha_X)$ values is better appreciated in Fig.~\ref{fig:epsilon_aX} for three values of $\Hinf$. All regions under the lines are excluded. For $\Hinf=10^9$\,GeV, the relic density can match the present-day one provided that $M_X$ ranges between ${\simeq}10^{11}$ and ${\simeq}10^{14}$\,GeV, $\alpha_X$ is less than ${\simeq}0.094$, and the efficiency $\epsilon$ is larger than ${\simeq}30\%$; otherwise the PIDM scenario cannot hold to explain the (entire) DM content observed today in the universe. For larger values of $\Hinf$, the allowed range of $(\epsilon,\alpha_X)$ gets larger as well as the allowed range of $M_X$. Larger values of $\alpha_X$ are possible on the condition of having $\epsilon$ larger than ${\simeq}~0.13\%$ (2.7\%) for $\Hinf=10^{13}$\,GeV ($10^{11}$\,GeV). However, note that the available parameter space shrinks significantly by restricting the allowed mass range to high values. For the mass of the PIDM particles to lie above $M_\text{GUT}$ for instance, the allowed range of $(\epsilon,\alpha_X)$ values then becomes $(\geq 0.30, \leq 0.087)$ for $\Hinf=10^{13}$\,GeV. Probing such a value of $\Hinf$ will be possible with the increased sensitivity to the tensor-to-scalar ratio through the B-mode polarisation of the cosmic microwave background anisotropies in the next decade~\cite{CMB-S4:2016ple,Ishino:2016wxl}. In parallel, the sensitivity to UHE photons will also improve thanks to the planned UHECR observatories~\cite{POEMMA:2020ykm,Horandel:2021prj}. Hence the parameter space allowing for GUT scale masses will be explored and could be either uncovered or significantly shrunk.

\section{Constraints from SM stability during inflation}
\label{sec:xi-aX_constraints}

As previously stated, the SM Higgs potential develops an instability at large field value. As a consequence, the SM electroweak vacuum does not correspond to minimum energy, but to a metastable state. Still, a quantitative estimation of its rate of quantum tunnelling into a lower energy state in flat space-time leads to a lifetime comfortably larger than the age of the universe~\cite{Buttazzo:2013uya,Andreassen:2017rzq}.
Such an astronomically-long lifetime is not challenged in the cosmological context due to thermal fluctuations allowing the decay when the temperature was high enough~\cite{Espinosa:2007qp}. Yet, it might be challenged due to large fluctuations of free fields generated by the dynamics on a curved background because of the presence of a non-minimal coupling $\xi$ between the Higgs field and the curvature of space-time during the inflation period. In such a case, new degrees of freedom at an intermediate scale below $\LambdaI$ would be necessary to stabilize the SM and the PIDM scenario would somehow be invalidated. 

\begin{figure}
\centering
\includegraphics[width=\columnwidth]{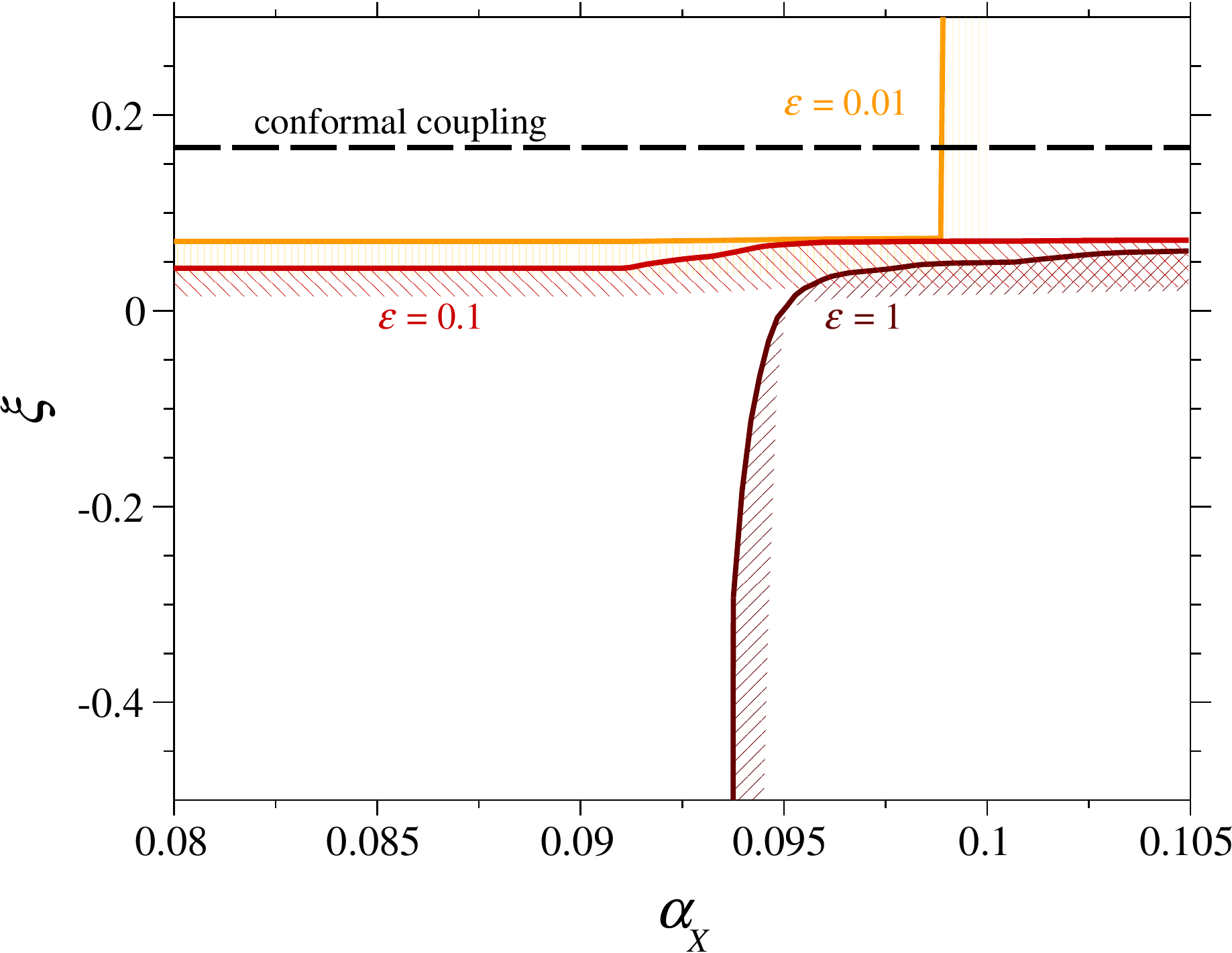}
\caption{Constraints from vacuum stability in the PIDM scenario. The excluded ranges of $(\xi,\alpha_X)$ values for the scenario of Planckian-interactive massive particles as DM is delineated for three examples of $\epsilon$ -- see text for details. For reference, the non-minimal value for $\xi$ expected from conformal theories is shown as the dashed line.}
\label{fig:xi_aX}
\end{figure}

The size of the field fluctuations aforementioned are critically determined by the Hubble rate parameter $\Hinf$, which governs the dynamics of the SM Higgs during inflation. The requirement for the electroweak vacuum to maintain its astronomical lifetime allows constraints between the non-minimal coupling $\xi$ and the Hubble rate $\Hinf$ in viable regions. Stability bounds have been derived in the $(\xi,\Hinf)$ plane by accounting for the curvature-dependent effective potential of the Higgs up to one-loop order~\cite{Markkanen:2018bfx}. They can be propagated into the $(\xi,\alpha_X)$ plane. To do so, a scan in the variable $\alpha_X$ is performed. For each value of $\alpha_X$, the corresponding upper limit on $M_X$ as obtained from Eq.~\eqref{eqn:tauX} is used in Eq.~\eqref{eqn:abundance} to determine the viable $\Hinf$ value, which is finally used to read the associated range of allowed values for $\xi$ from~\cite{Markkanen:2018bfx}. We show the result of the analysis in Fig.~\ref{fig:xi_aX} for three different values of efficiency. For $\epsilon=1$, the lower-right region delineated by the black curve is excluded. For $\epsilon=0.1$, the exclusion zone delineated by the red curve is enlarged. Finally, for $\epsilon=0.01$, the exclusion zone is delineated by the yellow curve: there are no possible values in the region $(\xi\lesssim 0.07,\alpha_X\gtrsim 0.099)$ for the PIDM scenario to hold. 

For reference, the value of $\xi=1/6$ that corresponds to a conformally-invariant coupling is shown as the dashed line. The experimental bounds from the LHC are $|\xi|\lesssim 2.6{\times}10^{15}$~\cite{Calmet:2017hja}.

\section{Conclusion}
\label{sec:conclusion}

In this paper, we have considered a class of SHDM scenarios in which the DM lifetime is stabilized due to having no charges under SM interactions.  In this case, DM may interact with SM particles through instantons of a gauge group describing the dark sector or only gravitationally.  We obtained constraints on the masses and couplings in such PIDM scenarios by exploiting the limits placed on the flux of UHE photons using the data of the Pierre Auger Observatory. In this case, super-heavy particles with masses as large as the GUT energy scale could be sufficiently abundant to match the DM relic density, provided that the inflationary energy scale is high ($\Hinf \sim 10^{13}$\,GeV) and the reheating efficiency is high (so that reheating is quasi-instantaneous). This rules out values of the dark-sector gauge coupling greater than ${\simeq}~0.09$. The mass values could however be smaller, relaxing the constraints on the efficiency. For more moderate values of $\epsilon$, the need to avoid more than one bubble nucleation event in the observable universe during inflation implies then that the non-minimal coupling of the Higgs to the curvature is more than a few percent. It is likely that the examples of constraints inferred on models of dark sectors and physics in the reheating epoch in the framework of inflationary cosmologies only scratch the surface of the power of limits on UHE photon fluxes to constrain physics otherwise beyond the reach of laboratory experiments. \\

\textit{Acknowledgments.} The successful installation, commissioning, and operation of the Pierre Auger Observatory would not have been possible without the strong commitment and effort from the technical and administrative staff in Malarg\"ue. We are very grateful to the following agencies and organizations for financial support:
Argentina -- Comisi\'on Nacional de Energ\'\i{}a At\'omica; Agencia Nacional de
Promoci\'on Cient\'\i{}fica y Tecnol\'ogica (ANPCyT); Consejo Nacional de
Investigaciones Cient\'\i{}ficas y T\'ecnicas (CONICET); Gobierno de la
Provincia de Mendoza; Municipalidad de Malarg\"ue; NDM Holdings and Valle
Las Le\~nas; in gratitude for their continuing cooperation over land
access; Australia -- the Australian Research Council; Belgium -- Fonds
de la Recherche Scientifique (FNRS); Research Foundation Flanders (FWO);
Brazil -- Conselho Nacional de Desenvolvimento Cient\'\i{}fico e Tecnol\'ogico
(CNPq); Financiadora de Estudos e Projetos (FINEP); Funda\c{c}\~ao de Amparo \`a
Pesquisa do Estado de Rio de Janeiro (FAPERJ); S\~ao Paulo Research
Foundation (FAPESP) Grants No.~2019/10151-2, No.~2010/07359-6 and
No.~1999/05404-3; Minist\'erio da Ci\^encia, Tecnologia, Inova\c{c}\~oes e
Comunica\c{c}\~oes (MCTIC); Czech Republic -- Grant No.~MSMT CR LTT18004,
LM2015038, LM2018102, CZ.02.1.01/0.0/0.0/16{\textunderscore}013/0001402,
CZ.02.1.01/0.0/0.0/18{\textunderscore}046/0016010 and
CZ.02.1.01/0.0/0.0/17{\textunderscore}049/0008422; France -- Centre de Calcul
IN2P3/CNRS; Centre National de la Recherche Scientifique (CNRS); Conseil
R\'egional Ile-de-France; D\'epartement Physique Nucl\'eaire et Corpusculaire
(PNC-IN2P3/CNRS); D\'epartement Sciences de l'Univers (SDU-INSU/CNRS);
Institut Lagrange de Paris (ILP) Grant No.~LABEX ANR-10-LABX-63 within
the Investissements d'Avenir Programme Grant No.~ANR-11-IDEX-0004-02;
Germany -- Bundesministerium f\"ur Bildung und Forschung (BMBF); Deutsche
Forschungsgemeinschaft (DFG); Finanzministerium Baden-W\"urttemberg;
Helmholtz Alliance for Astroparticle Physics (HAP);
Helmholtz-Gemeinschaft Deutscher Forschungszentren (HGF); Ministerium
f\"ur Innovation, Wissenschaft und Forschung des Landes
Nordrhein-Westfalen; Ministerium f\"ur Wissenschaft, Forschung und Kunst
des Landes Baden-W\"urttemberg; Italy -- Istituto Nazionale di Fisica
Nucleare (INFN); Istituto Nazionale di Astrofisica (INAF); Ministero
dell'Istruzione, dell'Universit\'a e della Ricerca (MIUR); CETEMPS Center
of Excellence; Ministero degli Affari Esteri (MAE); M\'exico -- Consejo
Nacional de Ciencia y Tecnolog\'\i{}a (CONACYT) No.~167733; Universidad
Nacional Aut\'onoma de M\'exico (UNAM); PAPIIT DGAPA-UNAM; The Netherlands
-- Ministry of Education, Culture and Science; Netherlands Organisation
for Scientific Research (NWO); Dutch national e-infrastructure with the
support of SURF Cooperative; Poland -- Ministry of Education and
Science, grant No.~DIR/WK/2018/11; National Science Centre, Grants
No.~2016/22/M/ST9/00198, 2016/23/B/ST9/01635, and 2020/39/B/ST9/01398;
Portugal -- Portuguese national funds and FEDER funds within Programa
Operacional Factores de Competitividade through Funda\c{c}\~ao para a Ci\^encia
e a Tecnologia (COMPETE); Romania -- Ministry of Research, Innovation
and Digitization, CNCS/CCCDI -- UEFISCDI, projects PN19150201/16N/2019,
PN1906010, TE128 and PED289, within PNCDI III; Slovenia -- Slovenian
Research Agency, grants P1-0031, P1-0385, I0-0033, N1-0111; Spain --
Ministerio de Econom\'\i{}a, Industria y Competitividad (FPA2017-85114-P and
PID2019-104676GB-C32), Xunta de Galicia (ED431C 2017/07), Junta de
Andaluc\'\i{}a (SOMM17/6104/UGR, P18-FR-4314) Feder Funds, RENATA Red
Nacional Tem\'atica de Astropart\'\i{}culas (FPA2015-68783-REDT) and Mar\'\i{}a de
Maeztu Unit of Excellence (MDM-2016-0692); USA -- Department of Energy,
Contracts No.~DE-AC02-07CH11359, No.~DE-FR02-04ER41300,
No.~DE-FG02-99ER41107 and No.~DE-SC0011689; National Science Foundation,
Grant No.~0450696; The Grainger Foundation; Marie Curie-IRSES/EPLANET;
European Particle Physics Latin American Network; and UNESCO.

We acknowledge for this work the support of the Institut Pascal at Universit\'e Paris-Saclay during the Paris-Saclay Astroparticle Symposium 2021, with the support of the P2IO Laboratory of Excellence (program ``Investissements d’avenir'' ANR-11-IDEX-0003-01 Paris-Saclay and ANR-10-LABX-0038), the P2I axis of the Graduate School Physics of Universit\'e Paris-Saclay, as well as IJCLab, CEA, IPhT, APPEC, the IN2P3 master projet UCMN and EuCAPT ANR-11-IDEX-0003-01 Paris-Saclay and ANR-10-LABX-0038).

\bibliographystyle{apsrev4-1}
\bibliography{biblio}

\end{document}